\documentclass[a4paper]{article}

\usepackage{INTERSPEECH2021}
\usepackage{xcolor}
\title{MBI-Net: A Non-Intrusive Multi-Branched Speech Intelligibility Prediction Model for Hearing Aids}
\name{Ryandhimas E. Zezario$^1$$^2$, Fei Chen$^3$, Chiou-Shann Fuh$^1$, Hsin-Min Wang$^2$, Yu Tsao$^2$}
\address{
  $^1$National Taiwan University\\
  $^2$Academia Sinica \\
  $^3$Southern University of Science and Technology of China
  }
\email{\{ryandhimas, yu.tsao\}@citi.sinica.edu.tw}

\begin{document}

\maketitle

\begin{abstract}
Improving the user's hearing ability to understand speech in noisy environments is critical to the development of hearing aid (HA) devices. For this, it is important to derive a metric that can fairly predict speech intelligibility for HA users. A straightforward approach is to conduct a subjective listening test and use the test results as an evaluation metric. However, conducting large-scale listening tests is time-consuming and expensive. Therefore, several evaluation metrics were derived as surrogates for subjective listening test results. In this study, we propose a multi-branched speech intelligibility prediction model (MBI-Net), for predicting the subjective intelligibility scores of HA users. MBI-Net consists of two branches of models, with each branch consisting of a hearing loss model, a cross-domain feature extraction module, and a speech intelligibility prediction model, to process speech signals from one channel. 
The outputs of the two branches are fused through a linear layer to obtain predicted speech intelligibility scores. Experimental results confirm the effectiveness of MBI-Net, which produces higher prediction scores than the baseline system in Track 1 and Track 2 on the Clarity Prediction Challenge 2022 dataset.

\end{abstract}

%In each of branch, the speech utterance is first processed by MSBG hearing loss to simulate speech utterance based on the hearing aid's listener characteristic. The generated hearing loss output will be used as the input to generate a cross-domain feature (the combination of spectral and end-to-end features along with the latent representation from self supervised learning model). The 
% 
\section{Introduction}
A fair way to assess speech intelligibility is critical for a variety of speech-related applications. Generally speaking, speech intelligibility measures refer to the ratio between the numbers of the correctly recognized words and the total words in the speech utterance. The most direct measure of speech intelligibility is the subjective listening test. However, conducting large-scale hearing tests is prohibitive. Therefore, a series of speech intelligibility measures based on signal processing have been proposed, such as \cite{ref_35}, speech intelligibility index (SII) \cite{ref_36}, extended SII (ESII) \cite{ref_37}, speech transmission index (STI) \cite{ref_38}, short-time objective intelligibility (STOI) \cite{ref_39}, and modified binaural short-time objective intelligibility (MBSTOI) \cite{ANDERSEN20181}. When calculating these metrics, a clean utterance is used as a reference to compare with the speech of interest. Thus, the usefulness is somewhat limited since a clean reference may not always be provided in real-world scenarios. To overcome this limitation, several non-intrusive speech intelligibility metrics have been developed, such as \cite{srmr, moda}, where a clean reference is not required. Compared to intrusive measures, non-intrusive speech intelligibility measures have better utility but tend to achieve lower performance.

%Specifically, developing a system that can produce a better intelligibility score in hearing aid is very important. Many hearing aid users struggle to recognize the word in a noisy environment accurately. However, performing speech intelligibility evaluation is very time-consuming and expensive, considering this method is based on the human listening tes

With the advent of deep learning (DL) models, several studies have used DL models to deploy non-intrusive speech intelligibility prediction models. Depending on the type of ground truth labels, the models can be divided into two different categories: the first category predicts objective speech intelligibility metrics, for example, STOI \cite{ref_55, ref_56, ref_52}. The second category aims to predict the subjective listening test results  \cite{ANDERSEN20181, SIP_DL}.

%When developing hearing aid (HA) devices, it is critical to check the user's ability to hear in different acoustic environments, thus requiring effective speech intelligibility measurements. 

Despite the remarkable performance achieved by the DL-based speech intelligibility prediction models, few studies have focused on designing speech intelligibility prediction models for HA users. A well-known system for estimating HA user's intelligibility scores is based on a combination of MBSTOI and sigmoid fitting  \cite{barker2022is}. Specifically, speech utterances are first processed by the MSBG model \cite{msbg, msbg2}, followed by MBSTOI score estimation. The estimated MBSTOI scores are then converted to the corresponding subjective intelligibility scores using a sigmoid fitting. Although this method achieves remarkable performance, the availability of clean utterances is a must for obtaining MBSTOI scores. Recently, a DL-based non-intrusive HA speech assessment network (HASA-Net) was proposed \cite{chiang2021hasa}. HASA-Net formulates the hearing loss pattern as a vector, which is combined with speech signals and then sent to a DL model to predict two HA evaluation metrics, namely the hearing aid speech quality index (HASQI) \cite{kates2014hearingb} and the hearing aid speech perception index (HASPI) \cite{kates2014hearinga}.

In our previous study, a multi-objective speech assessment model (MOSA-Net) \cite{mosa-net} was proposed to predict objective quality and intelligibility metrics for normal hearing individuals. MOSA-Net uses cross-domain features (spectral- and time-domain features, and latent representations from an SSL model \cite{hubert}) to obtain rich acoustic information; a convolutional neural network bidirectional long short-term memory architecture with a multiplicative attention mechanism (CNN-BLSTM+AT for short) is used as the main model architecture; a multi-task learning criterion (simultaneously predicting objective metric scores of quality, PESQ \cite{rix2001perceptual}, and intelligibility, STOI \cite{ref_39}), is employed to train the MOSA-Net model. Experimental results confirm that MOSA-Net can effectively predict PESQ and STOI scores, both of which yield very high correlations with ground truth scores. 

In this study, we extend MOSA-Net and develop a speech intelligibility prediction model for HA, called the multi-branched speech intelligibility prediction model (MBI-Net). MBI-Net consists two branches of models, and each branch contains a MSBG model, a cross-domain feature extraction module, and a speech intelligibility prediction model, to process speech signals from one channel. The outputs of the two branches are combined by a linear layer to predict speech intelligibility scores for HA users. Experimental results demonstrate that MBI-Net yields more accurate intelligibility prediction scores than the baseline system for both Track 1 and Track 2 on the Clarity Prediction Challenge 2022 dataset.

The rest of this paper is organized as follows. Section II reviews the related works of speech intelligibility prediction. Section III presents the proposed MBI-Net. Section IV describes experimental setup and results. Finally, Section V provides conclusions and future work.
\graphicspath{ {./images/} }
\begin{figure}[t]
\centering
\includegraphics[width=8cm]{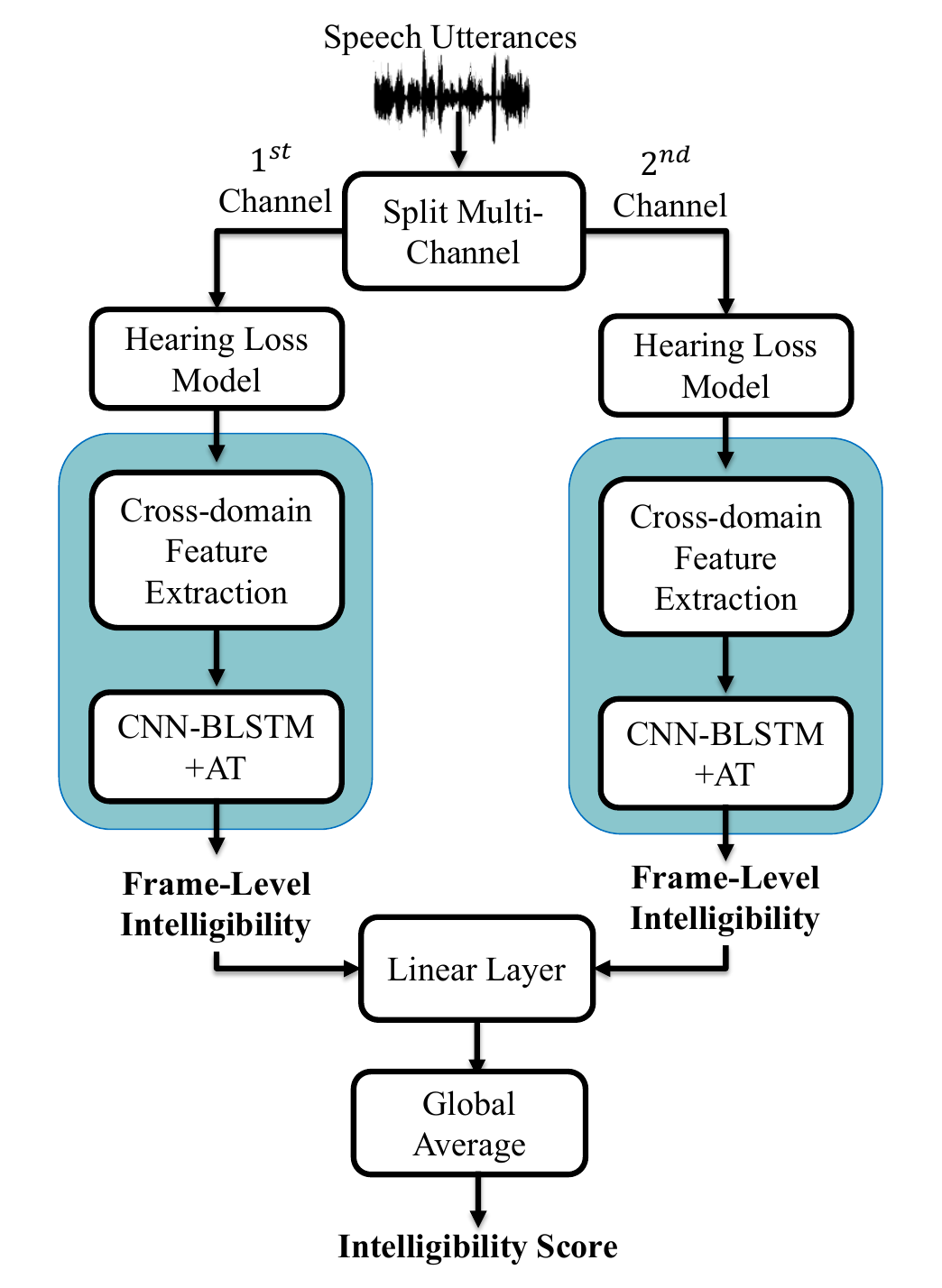} 
\caption{Architecture of the MBI-Net model.} 
\label{fig:MOSA-Net}
\end{figure}

\section{Related Works}
In recent years, DL-based models have been widely used as a base function for building speech intelligibility models. In \cite{ANDERSEN20181, SIP_DL}, a convolutional neural network (CNN) model is employed as the model architecture to process compressed spectral features and to estimate the subjective intelligibility score. In \cite{dong2020towards}, speech enhancement residuals were used to predict objective assessment score (PESQ and STOI). Furthermore, \cite{ref_52} proposed to use CNN-BLSTM+AT for predicting objective intelligibility score (STOI). In \cite{mosa-net}, a MOSA-Net, which uses cross-domain features, CNN-BLSTM+AT model architecture, and multi-task learning criterion, has been proposed to predict objective quality and intelligibility scores simultaneously; it also shows that the pretrained MOSA-Net can be easily adapted to a new model that can well predict subjective speech intelligibility scores using a small number of training samples. In the meanwhile, HASA-Net \cite{chiang2021hasa} was proposed to use a bidirectional long short-term memory (BLSTM) along with the multi-head attention mechanism to predict HASQI and HASPI scores simultaneously. Despite the remarkable performance achieved by DL-based speech intelligibility prediction models, few prior works have specifically estimated subjective/objective speech intelligibility scores for binaural HA users. Because the two ears of the user may have different hearing abilities, we should consider speech intelligibility separately and then combine them. This divide and conquer concept is similar to previous work using ensemble learning for speech signal processing \cite{deng2014ensemble, zhang2016deep, yu2020speech}. Along this idea, the proposed MBI-Net uses two branches of models, first processing the speech signal separately, and employing a fusion layer to compute the final prediction scores for binaural HA users.

\graphicspath{ {./images/} }
\begin{figure}[t]
\centering
\includegraphics[width=8cm]{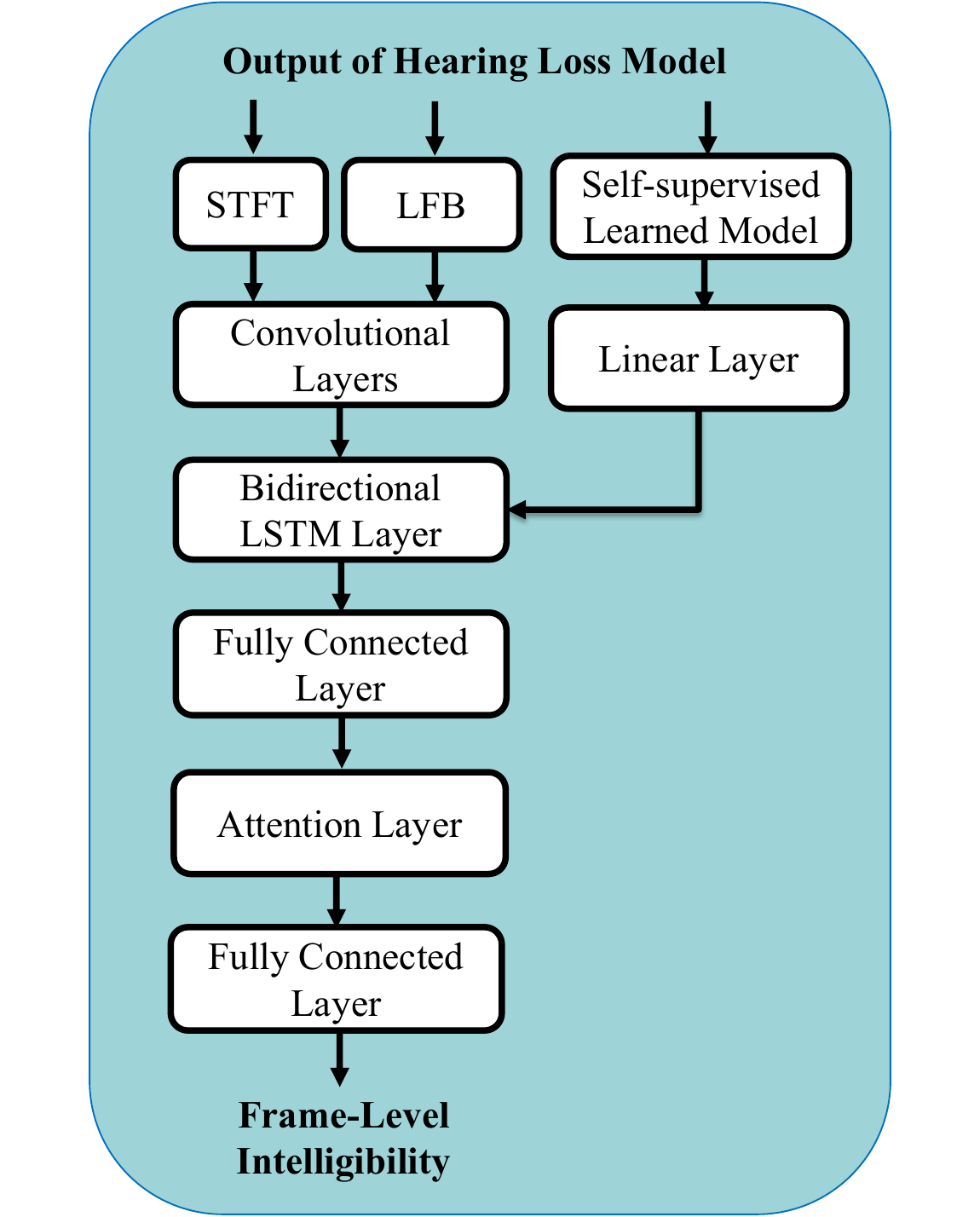} 
\caption{Illustration of extraction cross-domain feature and obtaining frame-level intelligibility score on CNN-BLSTM+AT architecture.} 
\label{fig:cross-domain}
\end{figure}

\section{MBI-Net}

The overall architecture of MBI-Net is shown in Fig. 1. As shown in the figure, MBI-Net consists of two branches of model, each characterizing one channel of speech signals in a binaural HA system. Specifically, given a dual-channel utterance, an audio splitting procedure is performed to divide the dual-channel speech signals into two monaural speech signals. The first and second channels correspond to the left and right ears, respectively. 
Each branch of MBI-Net consists of an MSBG model \cite{msbg, msbg2}, a cross-domain feature extraction module, and a frame-level speech intelligibility prediction model. 

The MSBG model modifies the speech signal according to the HA pattern and serves as a simulator to simulate the hearing ability of HA users. The modified speech signals are then sent to feature extraction and speech intelligibility prediction modules, as shown in Fig. 2. The feature extraction module consists of three parts:  (1) features obtained by converting speech waveforms through the short-time Fourier transform (STFT), termed spectral features; (2) features extracted by learnable filter banks (LFB) \cite{ravanelli2018speaker}, termed LFB features; (3) latent representations from self-supervised learning (SSL) models (HuBERT \cite{hubert} or WavLM \cite{chen2021wavlm}), termed SSL features. These three features are fed to the CNN-BLSTM with a multiplicative attention model to predict frame-level intelligibility scores. Finally, the predicted frame-level intelligibility scores of the two branches are fused by a linear layer along with global average pooling to obtain the final speech intelligibility score. To improve training stability, the objective function for training MBI-Net is a combined frame-level and utterance-level score, defined as follows:

\begin{equation}
\label{eq:loss}
   \small
    \begin{array}{c}
    O = \frac{1}{U}\sum\limits_{u=1}^U [(I_u-\hat{I}_u)^2+\frac{\alpha_{m}}{F_u}\sum\limits_{f=1}^{F_u}(I_u-\hat{i_{m_f}})^2] + \\ L_{left} + L_{right}
    \\
   L_{left}=\frac{\alpha_{l}}{F_u}\sum\limits_{f=1}^{F_u}(I_u-\hat{i_{l_f}})^2
   \\
   L_{right}=\frac{\alpha_{r}}{F_u}\sum\limits_{f=1}^{F_u}(I_u-\hat{i_{r_f}})^2   
    \\
    \end{array} 
\end{equation}
where $ L_{left}$ and $ L_{right}$, respectively, denote the frame-level loss of the left and right branches (ears); $\{I_u,\hat{I}_u\}$, respectively, denote the true and predicted utterance-level scores of the intelligibility; $U$ denotes the total number of training utterances; ${F}_u$ denotes the number of frames in the $u$-th training utterance; $\hat{i_{m_f}}$, $\hat{i_{l_f}}$, and $\hat{i_{r_f}}$ are the predicted frame-level scores of the intelligibility of the main branch, left branch, and right branch of the $f$-th frame, respectively; $\alpha_m$, $\alpha_l$, $\alpha_r$ are the weights between utterance-level and frame-level losses.

\section{Experiments}
In this section, we present the experimental setup and results of MBI-net on the Clarity Prediction Challenge 2022 dataset. 
\subsection{Experimental Setup}
The Clarity Prediction Challenge dataset 2022 included ten HA systems from the previous Clarity Enhancement Challenge 2021 \cite{clarity}. Twenty-five HA users participated in the listening test, and each listener was asked to answer what she/he heard from a played speech sample. The intelligibility score ranges from 0 to 100 (the higher the better). In addition, bilateral pure-tone audiograms of each listener were also estimated based on the hearing thresholds at [250, 500, 1000, 2000, 3000, 4000, 6000, 8000] Hz. For more detailed information, please refer to the the description paper provided by the Clarity Prediction Challenge organizer \cite{barker2022is}. The training set consisted of two tracks, Track 1 and Track 2. Track 1 consisted of 4863 training utterances, and Track 2 consisted of of 3580 training utterances.
The test set also consisted of tracks one and two with 2421 and 632 test utterances, respectively. For both tracks, the training and test utterances did not overlap. Three evaluation metrics, namely root mean square error (RMSE), standard deviation error (STDERR), and linear correlation coefficient (LCC), were used to evaluate the performance of MBI-Net. Lower RMSE and STDERR scores indicate that the predicted scores are closer to the ground-truth scores (lower is better). In contrast, a higher LCC score indicates that the predicted score has a higher correlation to the ground-truth score (higher is better).

\subsection{Experimental Result on Closed-Set}
\begin{table}[t]
\caption{RMSE, Standard Deviation, and LCC scores of Let-Branch, Right-Branch, MBI-Net (Ave), and MBI-Net (Lin) on the closed-set (Track 1) dataset.}
\footnotesize
\begin{center}
 \begin{tabular}{c||c||c||c} 
 \hline
 \hline
 \textbf{Systems} &\textbf{RMSE} & \textbf{STDERR} & \textbf{LCC}  \\ [0.5ex] 
 \hline
Left-Branch &25.33&0.51&0.73\\
Right-Branch &26.24&0.52&0.72 \\
MBI-Net (Ave) &25.12&0.51&0.74\\
MBI-Net (Lin) &\textbf{24.65}&\textbf{0.50}&\textbf{0.74}\\ 
 \hline

\end{tabular}
\end{center}
\end{table}

As mentioned in Section 3, the two-branches of models is designed to characterize the left and right ear audio channels, respectively. We first aimed to analyze the prediction capability of individual model, termed Left-Branch and Right-Branch, trained by monaural speech signals from the left and right channels, respectively. Both models were implemented based on Fig. 2, and an additional global average pooling was applied to average the frame-level intelligibility scores to obtain the utterance-level prediction score. Next, we compared different fusion strategies on the outputs of the two branches. As shown in Fig. 1, MBI-Net uses a linear layer followed by an average pooling to obtain the final prediction scores. For comparison, we replaced the linear layer with an average function. Accordingly, the outputs of the two channels were averaged and then sent to the pooling layer. This system is termed MBI-Net (Ave). Table 1 shows the prediction results of Left-Branch, Right-Branch, MBI-Net (Ave), and MBI-Net (denoted as MBI-Net (Lin) for clarity). All of the four systems reported in Table 1 used Hubert \cite{hubert} to obtain the SSL features. Experimental results confirm that all of the four systems can achieve satisfactory speech prediction performance. Next, MBI-Net (Ave) and MBI-Net (Lin) perform better than Left-Branch and Right-Branch, confirming the advantages of fusing information from the two branches of prediction model. Furthermore, MBI-Net (Lin) outperforms MBI-Net (Ave), suggesting that learned linear layer can facilitate MBI-Net to achieve better performance.

Next, we compare the performance of the baseline system with the proposed MBI-Net with different SSL features. Specifically, we deployed three MBI-Net systems: MBI-Net (Hub), MBI-Net (WavLM), and MBI-Net (WavLM+). All three systems used cross-domain features (spectral- and time-domain features, and SSL features) and the CNN-BLSTM+AT model architecture. MBI-Net (Hub) used HuBERT to deploy the SSL feature, and MBI-Net (WavLM) used WavLM to extract the SSL feature. MBI-Net (WavLM+) was the optimized version of MBI-Net (WavLM), where better parameter tuning was selected during training the model.

From Table 1. we first note that all the variants of MBI-Net can achieve better performance than the baseline systems in all evaluation metrics. Next, the SSL feature generated by WavLM tends to achieve overall better performance than the SSL feature from HuBERT, which confirms the advantages of WavLM to generate more representative features than HuBERT for deploying speech intelligibility model. Furthermore, the overall best performance is obtained by MBI-Net (WavLM+). For Clarity Prediction Challenge, we submitted the results obtained by MBI-Net (Hub) and MBI-Net (WavLM). After the challenge period, we obtained better results from MBI-Net (WavLM+). 

\begin{table}[t]
\caption{RMSE, Standard Deviation, and LCC scores of Baseline, MBI-Net (Hub), and MBI-Net (WavLM) on the closed-set (Track 1) dataset.}
\footnotesize
\begin{center}
 \begin{tabular}{c||c||c||c} 
 \hline
 \hline
 \textbf{Systems} &\textbf{RMSE} & \textbf{STDERR} & \textbf{LCC}  \\ [0.5ex] 
 \hline
Baseline&28.52&0.58&0.62 \\ \hline
MBI-Net (Hub)&24.65&0.50&0.74 \\ \hline
MBI-Net (WavLM)&24.06&0.49&0.75 \\ \hline
MBI-Net (WavLM+)&\textbf{23.05}&\textbf{0.46}&\textbf{0.78} \\ \hline
 \hline

\end{tabular}
\end{center}
\end{table}

\subsection{Experimental Result on Open-Set}
In the second set of experiments, we aimed to evaluate MBI-Net on a smaller training set, namely Track 2. Similar to the previous experiment, we intended to compare the performance of the baseline system with three versions of MBI-Net. All the systems used the identical configuration as the previous experiment. Experimental results confirmed that all the proposed MBI-Net could achieve overall better performance than the baseline systems. Interestingly, MBI-Net (WavLM) could significantly improve the baseline systems in all evaluation metrics. Moreover, the optimized version of MBI-Net (WavLM+) could further boost the performance for all evaluation metrics compared to MBI-Net (WavLM). These results again confirm the advantages of WavLM to deploying more representative features than HuBERT. In addition, these results confirm the benefit of a multi-branched model and cross-domain features for deploying speech intelligibility prediction models for HA.

\begin{table}[t]
\caption{RMSE, Standard Deviation, and LCC scores of Baseline, MBI-Net (Hub), and MBI-Net (WavLM) on the open-set (Track 2) dataset.}
\footnotesize
\begin{center}
 \begin{tabular}{c||c||c||c} 
 \hline
 \hline
 \textbf{Systems} &\textbf{RMSE} & \textbf{STDERR} & \textbf{LCC}  \\ [0.5ex]  \hline
Baseline&36.52&1.35&0.53\\ \hline
MBI-Net (Hub)&30.72&1.22&0.59 \\ \hline
MBI-Net (WavLM)&28.90&1.09&0.65 \\ \hline
MBI-Net (WavLM+)&\textbf{24.36}&\textbf{0.96}&\textbf{0.75} \\ \hline
 \hline

\end{tabular}
\end{center}
\end{table}

\subsection{Qualitative Analysis}

We also present the scatter plots of predicted targets of MBI-Net (WavLM+) and Baseline in Fig. 3. We only select MBI-Net (WavLM+) for qualitative analysis because MBI-Net (WavLM+) achieves the best performance among the other MBI-Net systems. From Fig. 3, MBI-Net (WavLM+) distributes more diagonally than baseline system, echoing the better results achieved by MBI-Net (WavLM+) shown in Tables 1-3. The results from Tables 1-3 and Fig. 3 confirm the advantages of deploying a multi-branched model and the capability of cross-domain features to enrich the acoustic information for obtaining an accurate speech intelligibility model for binaural HA users.

\graphicspath{ {./images/} }
\begin{figure}[t]
\centering
\includegraphics[width=8cm]{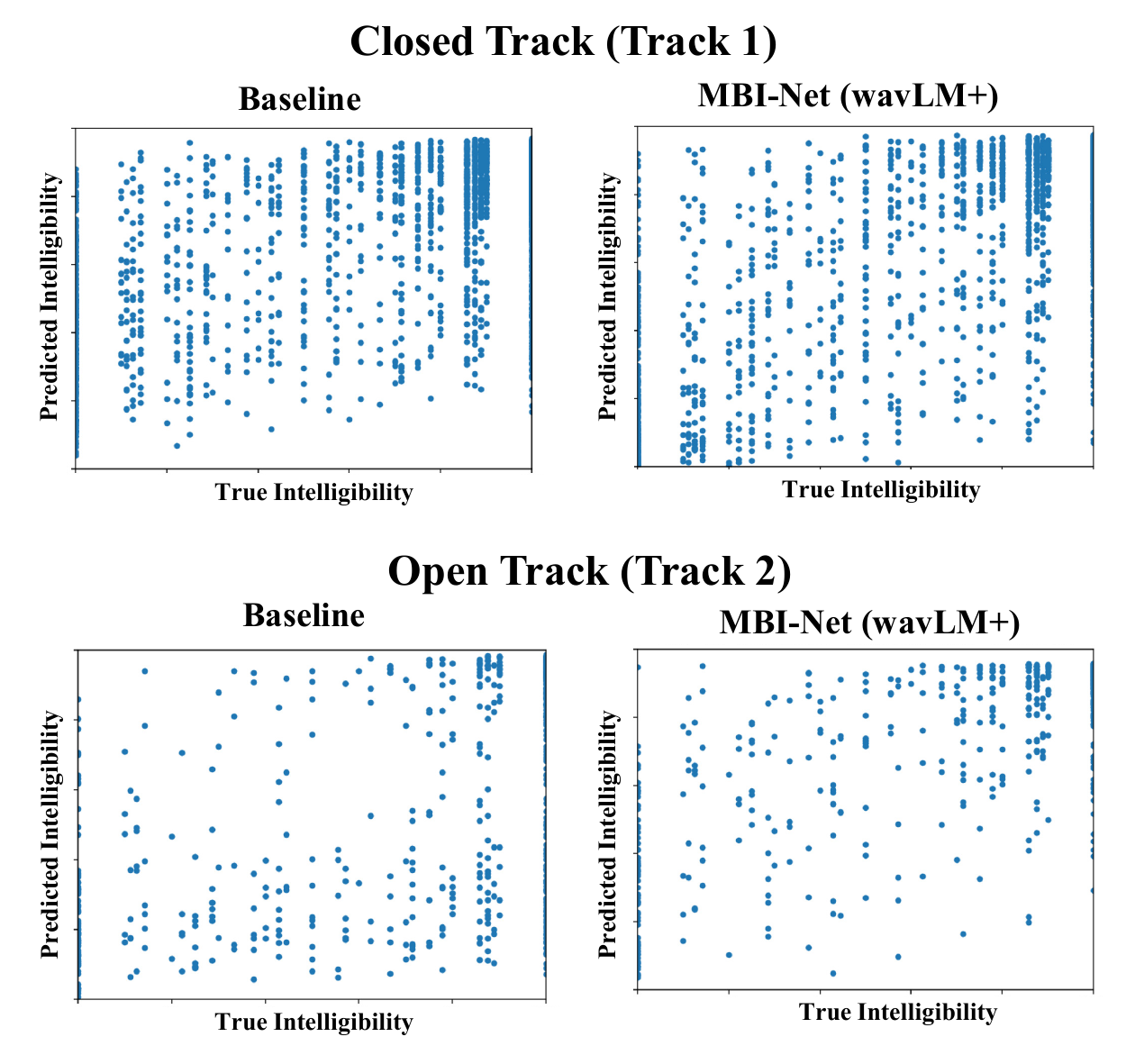} 
\caption{Scatter plots of two speech intelligibility prediction models: Baseline and MBI-Net (WavLM+).} 
\label{fig:scat}
\end{figure}

\section{Conclusion}
In this study, we presented MBI-Net, a multi-branched speech intelligibility prediction model for binaural HA users. MBI-Net adopts two-branches of models corresponding to two speech channels of the binaural HAs. Each branch of MBI-Net consists of an MSBG model, a cross-domain feature extraction module, and the CNN-BLSTM+AT model architecture. The outputs of the two branches are then fused through a linear layer to obtain the final speech intelligibility score. Experimental results from both Track 1 and Track 2 have confirmed the advantages of implementing the multi-branched model and using cross-domain features for achieving a better intelligibility prediction score. Furthermore, experimental results confirm the advantages of WavLM in deploying representative SSL features. In the future, we aim to explore and derive advanced cross-domain features and model architectures to further improve the model's speech intelligibility prediction performance for binaural HA users.

\bibliographystyle{IEEEtran}
\bibliography{mybib}

\begin{thebibliography}{10}
\providecommand{\url}[1]{#1}
\csname url@samestyle\endcsname
\providecommand{\newblock}{\relax}
\providecommand{\bibinfo}[2]{#2}
\providecommand{\BIBentrySTDinterwordspacing}{\spaceskip=0pt\relax}
\providecommand{\BIBentryALTinterwordstretchfactor}{4}
\providecommand{\BIBentryALTinterwordspacing}{\spaceskip=\fontdimen2\font plus
\BIBentryALTinterwordstretchfactor\fontdimen3\font minus
  \fontdimen4\font\relax}
\providecommand{\BIBforeignlanguage}[2]{{%
\expandafter\ifx\csname l@#1\endcsname\relax
\typeout{** WARNING: IEEEtran.bst: No hyphenation pattern has been}%
\typeout{** loaded for the language `#1'. Using the pattern for}%
\typeout{** the default language instead.}%
\else
\language=\csname l@#1\endcsname
\fi
#2}}
\providecommand{\BIBdecl}{\relax}
\BIBdecl

\bibitem{ref_35}
N.~R. French and J.~C. Steinberg, ``Factors governing the intelligibility of
  speech sounds,'' \emph{Journal of the Acoustical Society of America},
  vol.~19, no.~1, pp. 90--119, 1947.

\bibitem{ref_36}
A.~S.~S. 1997, ``Methods for calculation of the speech intelligibility index,''
  in \emph{Acoustical Society of America}, 1997.

\bibitem{ref_37}
T.~Houtgast and H.~.~M. Steeneken, ``Evaluation of speech transmission channels
  by using artificial signals,'' \emph{Acustica}, vol.~25, no.~6, pp. 355--367,
  1971.

\bibitem{ref_38}
H.~J.~M. Steeneken and T.~Houtgast, ``A physical method for measuring
  speech-transmission quality,'' \emph{Journal of the Acoustical Society of
  America}, vol.~67, no.~1, pp. 318--326, 1980.

\bibitem{ref_39}
C.~H. Taal, R.~C. Hendriks, R.~Heusdens, and J.~Jensen, ``An algorithm for
  intelligibility prediction of time-frequency weighted noisy speech,''
  \emph{IEEE/ACM Transactions on Audio, Speech and Language Processing},
  vol.~19, no.~7, pp. 2125--2136, 2011.

\bibitem{ANDERSEN20181}
A.~H. Andersenan, J.~M. Haan, Z.-H. Tan, and J.Jensen, ``Refinement and
  validation of the binaural short time objective intelligibility measure for
  spatially diverse conditions,'' \emph{Speech Communication}, vol. 102, pp.
  1--13, 2018.

\bibitem{srmr}
T.~H. Falk, C.~Zheng, and W.~Chan, ``A non-intrusive quality and
  intelligibility measure of reverberant and dereverberated speech,''
  \emph{IEEE Transactions on Audio, Speech, and Language Processing}, vol.~18,
  no.~7, pp. 1766--1774, 2010.

\bibitem{moda}
F.~Chen, O.~Hazrati, and P.~C. Loizou, ``Predicting the intelligibility of
  reverberant speech for cochlear implant listeners with a non-intrusive
  intelligibility measure,'' \emph{Biomedical Signal Processing and Control},
  vol.~8, no.~3, pp. 311--314, 2012.

\bibitem{ref_55}
X.~Jia and D.~Li, ``A deep learning-based time-domain approach for
  non-intrusive speech quality assessment,'' in \emph{Proc. APSIPA ASC}, 2020,
  pp. 477--481.

\bibitem{ref_56}
X.~Dong and D.~S. Williamson, ``An attention enhanced multi-task model for
  objective speech assessment in real-world environments,'' in \emph{Proc.
  ICASSP}, 2020, pp. 911--915.

\bibitem{ref_52}
R.~E. Zezario, S.-W. Fu, C.-S. Fuh, Y.~Tsao, and H.-M. Wang, ``{STOI-Net}: A
  deep learning based non-intrusive speech intelligibility assessment model,''
  in \emph{Proc. APSIPA ASC}, 2020, pp. 482--486.

\bibitem{SIP_DL}
M.~B. Pedersen, A.~H. Andersen, S.~H. Jensen, and J.~Jensen, ``A neural network
  for monaural intrusive speech intelligibility prediction,'' in \emph{Proc.
  ICASSP}, 2020, pp. 336--340.

\bibitem{barker2022is}
J.~Barker, M.~Akeroyd, T.~J. Cox, J.~F. Culling, J.~Firth, S.~Graetzer,
  H.~Griffiths, L.~Harris, G.~Naylor, Z.~Podwinska, E.~Porter, and R.~V. Munoz,
  ``The 1st clarity prediction challenge: A machine learning challenge for
  hearing aid intelligibility prediction,'' in \emph{Proc. INTERSPEECH}, 2022.

\bibitem{msbg}
T.~Baer and B.~C.~J. Moore, ``Effects of spectral smearing on the
  intelligibility of sentences in noise,'' \emph{The Journal of the Acoustical
  Society of America}, vol.~94, no.~3, pp. 1229--1241, 1993.

\bibitem{msbg2}
------, ``Effects of spectral smearing on the intelligibility of sentences in
  the presence of interfering speech,'' \emph{The Journal of the Acoustical
  Society of America}, vol.~95, no.~4, pp. 2277--2280, 1994.

\bibitem{chiang2021hasa}
H.-T. Chiang, Y.-C. Wu, C.~Yu, T.~Toda, H.-M. Wang, Y.-C. Hu, and Y.~Tsao,
  ``Hasa-net: A non-intrusive hearing-aid speech assessment network,'' in
  \emph{2021 IEEE Automatic Speech Recognition and Understanding Workshop
  (ASRU)}, 2021, pp. 907--913.

\bibitem{kates2014hearingb}
J.~M. Kates and K.~H. Arehart, ``The hearing-aid speech quality index (hasqi)
  version 2,'' \emph{Journal of the Audio Engineering Society}, vol.~62, no.~3,
  pp. 99--117, 2014.

\bibitem{kates2014hearinga}
------, ``The hearing-aid speech perception index (haspi),'' \emph{Speech
  Communication}, vol.~65, pp. 75--93, 2014.

\bibitem{mosa-net}
R.~E. Zezario, S.-W. Fu, F.~Chen, C.-S. Fuh, H.-M. Wang, and Y.~Tsao, ``Deep
  learning-based non-intrusive multi-objective speech assessment model with
  cross-domain features,'' \emph{arXiv:2110.02635}, 2022.

\bibitem{hubert}
W.-N. Hsu, B.~Bolte, Y.-H.~H. Tsai, K.~Lakhotia, R.~Salakhutdinov, and
  A.~Mohamed, ``{HuBERT}: Self-supervised speech representation learning by
  masked prediction of hidden units,'' \emph{IEEE/ACM Transactions on Audio,
  Speech, and Language Processing}, 2021.

\bibitem{rix2001perceptual}
A.~W. Rix, J.~G. Beerends, M.~P. Hollier, and A.~P. Hekstra, ``Perceptual
  evaluation of speech quality (pesq)-a new method for speech quality
  assessment of telephone networks and codecs,'' in \emph{2001 IEEE
  international conference on acoustics, speech, and signal processing.
  Proceedings (Cat. No. 01CH37221)}, vol.~2.\hskip 1em plus 0.5em minus
  0.4em\relax IEEE, 2001, pp. 749--752.

\bibitem{dong2020towards}
X.~Dong and D.~S. Williamson, ``Towards real-world objective speech quality and
  intelligibility assessment using speech-enhancement residuals and
  convolutional long short-term memory networks,'' \emph{The Journal of the
  Acoustical Society of America}, vol. 148, no.~5, pp. 3348--3359, 2020.

\bibitem{deng2014ensemble}
L.~Deng and J.~Platt, ``Ensemble deep learning for speech recognition,'' in
  \emph{Proc. INTERSPEECH}, 2014, pp. 1915--1919.

\bibitem{zhang2016deep}
X.-L. Zhang and D.~Wang, ``A deep ensemble learning method for monaural speech
  separation,'' \emph{IEEE/ACM transactions on audio, speech, and language
  processing}, vol.~24, no.~5, pp. 967--977, 2016.

\bibitem{yu2020speech}
C.~Yu, R.~E. Zezario, S.-S. Wang, J.~Sherman, Y.-Y. Hsieh, X.~Lu, H.-M. Wang,
  and Y.~Tsao, ``Speech enhancement based on denoising autoencoder with
  multi-branched encoders,'' \emph{IEEE/ACM Transactions on Audio, Speech, and
  Language Processing}, vol.~28, pp. 2756--2769, 2020.

\bibitem{ravanelli2018speaker}
M.~Ravanelli and Y.~Bengio, ``Speaker recognition from raw waveform with
  sincnet,'' in \emph{2018 IEEE Spoken Language Technology Workshop (SLT)},
  2018, pp. 1021--1028.

\bibitem{chen2021wavlm}
S.~Chen, C.~Wang, Z.~Chen, Y.~Wu, S.~Liu, Z.~Chen, J.~Li, N.~Kanda,
  T.~Yoshioka, X.~Xiao \emph{et~al.}, ``Wavlm: Large-scale self-supervised
  pre-training for full stack speech processing,'' \emph{arXiv preprint
  arXiv:2110.13900}, 2021.

\bibitem{clarity}
S.~Graetzer, J.~Barker, M.~A. T.~J.~Cox, J.~F. Culling, G.~Naylor, E.~Porter,
  and R.~V. Muñoz, ``Clarity-2021 challenges: Machine learning challenges for
  advancing hearing aid processing,'' in \emph{Proc. INTERSPEECH}, 2021, pp.
  686--690.

\end{thebibliography}

% \begin{thebibliography}{9}
% \bibitem[1]{Davis80-COP}
%   S.\ B.\ Davis and P.\ Mermelstein,
%   ``Comparison of parametric representation for monosyllabic word recognition in continuously spoken sentences,''
%   \textit{IEEE Transactions on Acoustics, Speech and Signal Processing}, vol.~28, no.~4, pp.~357--366, 1980.
% \bibitem[2]{Rabiner89-ATO}
%   L.\ R.\ Rabiner,
%   ``A tutorial on hidden Markov models and selected applications in speech recognition,''
%   \textit{Proceedings of the IEEE}, vol.~77, no.~2, pp.~257-286, 1989.
% \bibitem[3]{Hastie09-TEO}
%   T.\ Hastie, R.\ Tibshirani, and J.\ Friedman,
%   \textit{The Elements of Statistical Learning -- Data Mining, Inference, and Prediction}.
%   New York: Springer, 2009.
% \bibitem[4]{YourName17-XXX}
%   F.\ Lastname1, F.\ Lastname2, and F.\ Lastname3,
%   ``Title of your INTERSPEECH 2021 publication,''
%   in \textit{Interspeech 2021 -- 20\textsuperscript{th} Annual Conference of the International Speech Communication Association, September 15-19, Graz, Austria, Proceedings, Proceedings}, 2020, pp.~100--104.
% \end{thebibliography}

\end{document}